\documentclass[conference]{IEEEtran}
\IEEEoverridecommandlockouts
\usepackage{cite}
\usepackage{amsmath,amssymb,amsfonts}
\usepackage{algorithm}
\usepackage{algpseudocode}
\usepackage{graphicx}
\usepackage{textcomp}
\usepackage{xcolor}
\def\BibTeX{{\rm B\kern-.05em{\sc i\kern-.025em b}\kern-.08em
    T\kern-.1667em\lower.7ex\hbox{E}\kern-.125emX}}

\usepackage{caption}
\usepackage{subcaption}
\usepackage{multirow}
\usepackage{makecell}  
\usepackage[group-separator={,}]{siunitx}
\usepackage{hyperref}

\begin{document}


\title{Predicting the Output Structure of Sparse Matrix Multiplication with Sampled Compression Ratio}

\author{\IEEEauthorblockN{Zhaoyang~Du\IEEEauthorrefmark{1},
Yijin~Guan\IEEEauthorrefmark{2},
Tianchan~Guan\IEEEauthorrefmark{2}, 
Dimin~Niu\IEEEauthorrefmark{2},
Nianxiong~Tan\IEEEauthorrefmark{1},\\
Xiaopeng~Yu\IEEEauthorrefmark{1},
Hongzhong~Zheng\IEEEauthorrefmark{2},
Jianyi~Meng\IEEEauthorrefmark{2}, 
Xiaolang~Yan\IEEEauthorrefmark{1}, and
Yuan~Xie\IEEEauthorrefmark{2},~\IEEEmembership{Fellow,~IEEE}
}
\IEEEauthorblockA{\IEEEauthorrefmark{1}College of  Information  Science  and  Electronic Engineering, Zhejiang University, Hangzhou 310007, China}
\IEEEauthorblockA{\IEEEauthorrefmark{2}Alibaba Group, Hangzhou 311121, China}
\thanks{This work has been submitted to the IEEE for possible publication. Copyright may be transferred without notice, after which this version may no longer be accessible. Corresponding author: Zhaoyang~Du (email: 11731021@zju.edu.cn).}}

\maketitle

\begin{abstract}
Sparse general matrix multiplication (SpGEMM) is a fundamental building block in numerous scientific applications. One critical task of SpGEMM is to compute or predict the structure of the output matrix (i.e., the number of nonzero elements per output row) for efficient memory allocation and load balance, which impact the overall performance of SpGEMM. Existing work either precisely calculates the output structure or adopts upper-bound or sampling-based methods to predict the output structure. However, these methods either take much execution time or are not accurate enough. In this paper, we propose a novel sampling-based method with better accuracy and low costs compared to the existing sampling-based method. The proposed method first predicts the compression ratio of SpGEMM by leveraging the number of intermediate products (denoted as FLOP) and the number of nonzero elements (denoted as NNZ) of the same sampled result matrix. And then, the predicted output structure is obtained by dividing the FLOP per output row by the predicted compression ratio. We also propose a reference design of the existing sampling-based method with optimized computing overheads to demonstrate the better accuracy of the proposed method. We construct 625 test cases with various matrix dimensions and sparse structures to evaluate the prediction accuracy. Experimental results show that the absolute relative errors of the proposed method and the reference design are 1.56\% and 8.12\%, respectively, on average, and 25\% and 156\%, respectively, in the worst case.

\end{abstract}

\begin{IEEEkeywords}
Sparse matrix multiplication, SpGEMM, predicting output structure, nonzero structure, size estimation
\end{IEEEkeywords}

\section{Introduction}\label{sec:intro}
Sparse general matrix multiplication (SpGEMM) is a fundamental building block in numerous scientific and machining learning applications such as Markov clustering~\cite{markov}, algebraic multigrid solvers~\cite{AMG, AMG2}, molecular dynamics simulations~\cite{molecular}, multi-source breadth first search~\cite{BFS}, and finite element simulations based on domain decomposition~\cite{finite-element}.

Given two sparse matrices $A$ and $B$, SpGEMM computes the matrix multiplication $C = AB$, where $C$ is the sparse output matrix. To reduce the memory footprint and computation complexity when performing SpGEMM, all the input and output sparse matrices should be stored in a sparse matrix format. However, processing the sparse matrices with a sparse matrix format causes many performance issues due to the irregularities of the three involved matrices and the unknown structure of the output matrix.

Two essential aspects critical to SpGEMM's performance are the memory allocation method~\cite{yusuke,bhsparse} for the output matrix and the load balance method~\cite{bhash, nsparse, speck} when performing SpGEMM. Both performance issues are highly related to the output structure (i.e., the number of nonzero elements of each output row) of SpGEMM and how the output structure is computed~\cite{yusuke,bhsparse, bhash, nsparse, speck}.

Existing methods to compute the output structure include the precise-method~\cite{yusuke, nsparse, kokkos, speck}, the upper-bound method~\cite{yusuke, ESC, ESC2, bhsparse}, and the sampling-based method~\cite{better-size, bhash}. The precise method computes the exact output structure, which is called the symbolic phase, before performing the actual numeric matrix multiplication~\cite{nsparse}. The benefit is that the SpGEMM library adopting the precise method does not need to allocate the intermediate result matrix. However, the major problem with the precise method is that the computation complexity of the symbolic phase is similar to that of the numeric phase~\cite{yusuke, nsparse, speck}. As a result, the symbolic phase (precise method) takes non-trivial computing overheads. 

In contrast, the upper-bound method computes the number of necessary intermediate products (denoted as FLOP~\cite{yusuke}) per output row as the output structure, which is a low-cost computation step~\cite{yusuke, bhsparse}. However, the upper-bound method may allocate much memory space for the intermediate result matrix. For example, for a SpGEMM task with a compression ratio of 10, the memory footprint of the intermediate result matrix is 10$\times$ that of the actual result matrix. The compression ratio of a SpGEMM task (or simply of a result matrix) is defined by dividing the total FLOP to perform SpGEMM by the total number of nonzero elements (denoted as NNZ) of the result matrix.

Due to these limitations of the aforementioned methods, in this paper, we focus on developing a novel sampling-based method targeting high prediction accuracy and low computing cost.
The existing sampling-based method first randomly samples two sub-matrices of the two input matrices and computes the NNZ of the sampled result matrix (denoted as sampled NNZ). It then divides the sampled NNZ by $p$ to predict the NNZ of the result matrix, where $p$ represents the matrix size proportion of the sampled result matrix to the entire result matrix. 

To predict the output structure of SpGEMM, the existing sampling-based method has to compute the precise FLOP per output row and the total FLOP of the result matrix. And then, the predicted compression ratio of the result matrix is computed by dividing the total FLOP by the predicted total NNZ of the result matrix. At last, the output structure is predicted by dividing the FLOP per output row by the predicted compression ratio. Since the FLOP per output row and the total FLOP are precisely calculated, the accuracy of the predicted output structure, the predicted compression ratio, and the predicted NNZ of the result matrix can be seen as equivalent.

An intuition is that when the NNZ of a sampled result matrix is larger than its expectation, the FLOP of the same sampled result matrix (denoted as sampled FLOP) may also be larger than its expectation. Note that the expectations of the sampled NNZ or sampled FLOP are the precise NNZ($C$) or FLOP($C$) multiplied by $p$, where $p$ represents the proportion of the matrix size of the sampled result matrix to the entire result matrix. In other words, the relative errors of the sampled NNZ and sampled FLOP compared to their expectations may have a positive correlation. 

Based on the aforementioned intuition, we propose a novel sampling-based method, which exploits the potential positive correlation of the sampled NNZ and the sampled FLOP of the same sampled result matrix. The proposed method divides the sampled FLOP by the sampled NNZ to obtain a sampled compression ratio as the predicted compression ratio of the result matrix. By doing so, the prediction of the compression ratio may achieve a certain degree of error neutralization between the sampled FLOP and sampled NNZ. For example, suppose the relative errors of the sampled FLOP and the sampled NNZ are 25\%  and 30\%, respectively. In that case, the relative error of the predicted compression ratio will be 3.85\% (analyzed in Section~\ref{sec:proposed-method}). At last, the predicted output structure is easily computed by dividing the FLOP per output row by the predicted compression ratio.

As for the computing overheads, the existing sampling-based method uses the inner-product dataflow~\cite{bhash} to compute the sampled matrices, which is less efficient than the row-wise dataflow~\cite{brmerge, matraptor} on both the computing and sampling of the two input matrices. Therefore, in this work, we implement the proposed method using the row-wise dataflow.
To fairly compare the prediction accuracy of the proposed method and the existing method, we also implement a reference design of the existing sampling-based method using the row-wise dataflow and the same associated sampling method.

We select 25 representative real-world sparse matrices from the SuiteSparse~\cite{suitesparse} dataset and compose 625 test cases by multiplying them with each other to evaluate the prediction accuracy of the predicted NNZ($C$). The relative prediction errors of the proposed method and the reference design are 1.56\% and 8.12\% (smaller is better), respectively, on average, and 25\% and 156\%, respectively, in the worst case. We also conduct experiments to show that the parallel implementation of the proposed method only takes on average 0.78\% execution time of an entire state-of-the-art SpGEMM library (BRMerge-Precise)\cite{brmerge}.

The main contributions of this work are listed as follows:
\begin{itemize}
    \item We propose a novel sampling-based method that utilizes both the sampled FLOP and sampled NNZ to predict the output structure of SpGEMM.
    \item We propose a row-wise implementation of the proposed method and a reference design of the existing sampling-based method.
    \item We conduct comprehensive evaluations to show that the proposed method is much more accurate than the reference design of the existing method. We also conduct experiments to show that the parallel implementation of the proposed method only takes on average 0.78\% execution time of an entire state-of-the-art SpGEMM library (BRMerge-Precise)\cite{brmerge}.
\end{itemize}

The rest of this paper is organized as follows. Section~\ref{sec:bg} introduces the notations and backgrounds. Section~\ref{sec:relate} discusses the related work. Section~\ref{sec:method} describes the proposed prediction method and the reference design of the existing prediction method. Section~\ref{sec:impelmentation} details the efficient parallel implementation of the proposed method. Section~\ref{sec:experiment} shows the evaluation accuracy of the proposed method and the reference design. This section also shows the computing overheads of the proposed method. Section~\ref{sec:conclude} concludes this paper.

The source code of this paper is provided in \url{https://github.com/lorentzbf/Size-Prediction.git}.

\section{Preliminaries and Backgrounds}\label{sec:bg}
\subsection{Notations}
Table~\ref{tab:notation} defines the notations used in this paper. The matrix dimensions of the three matrices $A$, $B$, and $C$ are $M \times K$, $K \times N$, and $M \times N$, respectively.

\begin{table}[h]
    \centering
    \begin{tabular}{|l|l|}
        \hline
        Notation & Explanation \\
        \hline
        $A$ & The first input matrix\\
        \hline
        $B$ & The second input matrix \\
        \hline
        $C$ & The result matrix \\
        \hline
        NNZ($\cdot$) & Number of nonzero elements  \\
        \hline
        FLOP($\cdot$) & Number of intermediate products\\
        \hline
        CR & Compression ratio, defined as FLOP/NNZ \\
        \hline
    \end{tabular}
    \caption{Notations used in this paper.}
    \label{tab:notation}
\end{table}

\subsection{CSR Storage Format}\label{sec:CSR}
\begin{figure}[h]
\centering
\includegraphics[width=0.45\textwidth]{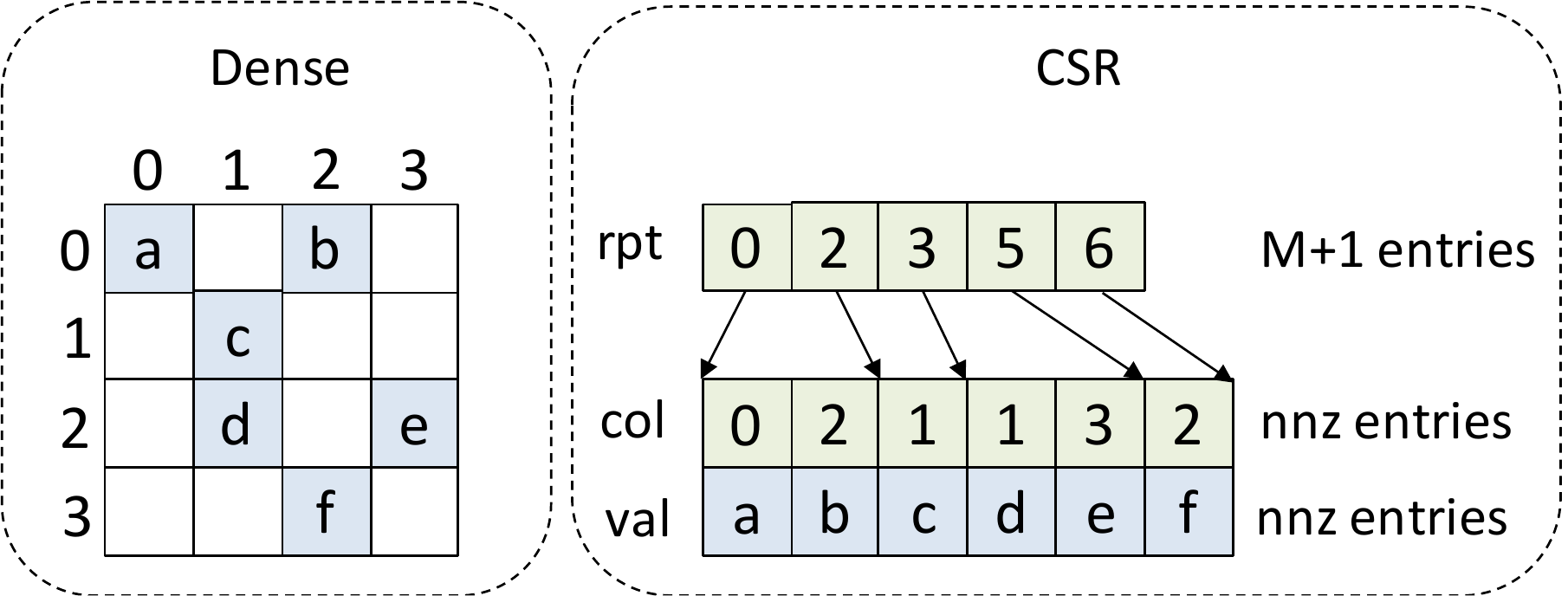}
\caption{Illustration of the CSR storage format. Left: dense storage format. Right: CSR storage format.}
\label{fig:CSR}
\end{figure}

CSR (Compressed Sparse Row) storage format stores the sparse matrix in a compressed way. Fig.~\ref{fig:CSR} illustrates the CSR storage format, which consists of three arrays named $rpt$, $col$, and $val$. The $val$ and $col$ arrays record the nonzero elements and their corresponding column indices in a sorted row-major and column-major order. The $rpt$ array records the start and end offsets for each row's values and column indices in the $val$ and $col$ arrays. 

\subsection{Row-wise Dataflow}\label{sec:row-wise}
The row-wise dataflow is described by \eqref{eq:row-wise}, which shows the computation of the $i_{th}$ output row ($C_{i*}$).

\begin{equation}C_{i*} = \sum_{k}A_{ik}\cdot{B}_{k*},\label{eq:row-wise}\end{equation}
where $C_{i*}$ and $A_{i*}$ represent all the nonzero elements in the $i_{th}$ row of $C$ and $A$, respectively, $k$ belongs to the set of column indices of $A_{i*}$, and $B_{k*}$ represents all the nonzero elements in the $k_{th}$ row of $B$.

\section{Related Work}\label{sec:relate}
We first describe a highly related prediction method proposed in the database literature. And then, we describe two sampling-based methods that predict the output structure of SpGEMM. At last, we describe several SpGEMM libraries that utilize the output structure for memory allocation and load balance issues.

Bar-Yossef~\textit{et al.}~\cite{distinct-stream} proposed a method to predict the number of distinct elements in a data stream.
Let $\boldsymbol{a} = a_{1}, a_{2}, ... a_{n}$ be an array with $n$ elements, where $a_{i} \in [0,m]$. Certain elements in $\boldsymbol{a}$ may have the same value. Let $F = F(\boldsymbol{a})$ represent the number of distinct elements in $\boldsymbol{a}$. To predict $F$, this method first constructs a hash function $h: [m] \rightarrow [0,1]$. And then, the method applies $h(\cdot)$ to \textbf{all} elements in $\boldsymbol{a}$ and maintain the smallest hashed value as $v$. At last, the predicted number of distinct elements in $\boldsymbol{a}$ is computed by $F^{*} = 1/v$. The intuition here is that if the hashed values of all the elements in $\boldsymbol{a}$ are randomly distributed in $[0,1]$, the smallest hashed value will be $1/F$. This prediction method is utilized in the following methods that predict the output structure of SpGEMM~\cite{better-size, bhash}. The benefit of this method is minimum memory space usage when processing the data in $\boldsymbol{a}$ since it only maintains the smallest element. However, the time complexity is not necessarily low since all the elements in $\boldsymbol{a}$ are processed.

Amossen~\textit{et al.}~\cite{better-size} proposed a method to predict the total NNZ of the result matrix of SpGEMM, which is mainly based on the idea in~\cite{distinct-stream}. This method constructs a hash function $h: [m,n] \rightarrow [0,1]$ and samples random subsets from the two input matrices as $A'$ and $B'$. It applies all the row and column indices of the intermediate products when computing $C'  = A'B'$ to the hash function $h$. The $k_{th}$ smallest element of the hashed values is maintained as $v$. After this process, the predicted NNZ($C'$) is computed by $k/v$ and the predicted NNZ($C$) is computed by $k/v/p$, where $p$ stands for the matrix size proportion of $C'$ compared to $C$.

Pham~\textit{et al.}~\cite{bhash} proposed a SpGEMM library BHash which implements Amossen's prediction method. Specifically, Pham's method randomly samples the rows from $A$ as $A'$ with the probability $p = 1/10$ and randomly samples columns from  $B$ as $B'$ with the same probability. It then use the same method in~\cite{better-size} to predict NNZ($C'$), where $C' = A'B'$. At last, the predicted NNZ($C$) is calculated by $\mathit{NNZ(C')}/p^2$. 
The predicted NNZ($C$) and the predicted output structure are used for the load balance of the computing tasks and memory allocation for the result matrix in BHash~\cite{bhash}.

Most of the existing SpGEMM libraries only use the precise or upper-bound methods to compute/predict the output structure~\cite{yusuke, cusparse, ESC, ESC2, bhsparse, rmerge, nsparse, speck}. For example, bhsparse~\cite{bhsparse} computes the FLOP per output row as the upper-bound output structure and classifies the rows with different FLOP to different bins. Rows in different bins are computed by different accumulation methods for better load balance. The memory allocation of bhsparse is also based on the upper-bound output structure. The nsparse~\cite{nsparse} computes the upper-bound output structure and uses this information for the load balance of its symbolic phase, which computes the precise output structure.

\section{Proposed Method}\label{sec:method}

This section describes the proposed sampling-based method and the reference design of the existing sampling-based method. The proposed method contains three progressive optimizations compared to the existing method: 1) adopting the row-wise dataflow with the associated sampling method, 2) calculating the precise NNZ of the sampled result matrix, 3) predicting the NNZ of the result matrix by using the sampled FLOP and sampled NNZ. To fairly compare the prediction accuracy of the proposed method and the existing method, we implement the reference design of the existing method by adopting the first two optimizations.

\subsection{Computing Dataflow and Sampling Method}
The existing method selects both rows and columns from the two input matrices~\cite{better-size,bhash}. However, the two input matrices are usually stored in homogeneous storage format~\cite{bhsparse, nsparse}. For example, both input matrices may be stored in the CSR format, which is hard to select the columns from the matrix. Furthermore, even if the rows and columns can be selected with a low cost, the following inner-product dataflow may still be a performance bottleneck~\cite{brmerge,matraptor}. To tackle these two performance issues, we adopt the row-wise dataflow for the sampling method and the computation of the samples.

The associated sampling method randomly samples $p$ fraction of rows from the $A$ matrix as $A'$, where $p$ is usually a small value for a low cost. The method does not need to sample the $B$ matrix since $B$ is accessed according to the selected rows of $A$ (see Section~\ref{sec:row-wise}). This can also be seen as the entire $B$ matrix is sampled. 

\subsection{Computing Method for the Samples}
The existing method does not compute the precise NNZ of the samples, but predicts it by utilizing a hash function $h: [m,n] \rightarrow [0,1]$~\cite{better-size,bhash}. The accuracy of the existing method is highly related to the hash function, which is difficult to be constructed. Moreover, the computation complexity of the existing method is similar to that of precisely computing the NNZ of the samples. The reason is that the bottlenecks of both methods lie in processing all the intermediate products of the samples. Therefore, we propose to directly computes the precise NNZ of the samples, which is automatically more accurate in computing the NNZ of the samples than the existing sampling-based method. We describe the parallel implementation of computing the precise NNZ of the samples in section~\ref{sec:compute-nnz}.

\subsection{Reference Design}\label{sec:reference}\label{sec:reference-design}
The reference design adopts the two techniques mentioned above. We then describe how the reference design predicts the output structure in more detail. We also analyze the relative error of the reference design.

We denote the computed NNZ of the samples as $z^*$, the expected NNZ of the samples as $z$, the predicted NNZ($C$) as $Z^*$, and the precise NNZ($C$) as $Z$. Note that throughout this paper, we use the additional subscript $\cdot_1$ to denote the variables computed by the reference design. The expected NNZ of the samples $z$ is defined as $z = pZ$, where $p$ means the proportion of the matrix size of the sampled result matrix compared to the entire result matrix. For example, if the number of rows of the $A$ matrix is $1000$ and the number of sampled rows from the $A$ matrix is 3, $p$ will be equal to $0.003$. Similar to the previous work~\cite{better-size, bhash}, the reference design predicts the NNZ($C$) by $z_1^*/p$. We define the relative error of the $z_1^*$ to the expected $z$ as $\epsilon_1$. Then the entire prediction method and the error analysis of the reference design are shown in~\eqref{eq:Z1_star}.

\begin{equation}
    Z_1^* = \frac{1}{p}z_1^* = \frac{1}{p}(1+\epsilon_1)z_1 = (1+\epsilon_1)Z.
    \label{eq:Z1_star}
\end{equation}

To predict the output structure of the result matrix, the reference design (as well as the existing methods~\cite{better-size,bhash}) has to compute the FLOP per output row and the total FLOP of the result matrix (denoted as $F$). And then, the predicted compression ratio is computed by $F/Z_1^*$, and the output structure is computed by dividing the FLOP per output row by the predicted compression ratio. 

\subsection{Proposed Predicting Method}\label{sec:proposed-method}
The proposed prediction method utilizes the same information computed by the reference design, which includes the computed FLOP per output row and the NNZ of the samples. In addition, the proposed method also computes the FLOP of the samples. The cost of computing the FLOP of the samples is negligible after the FLOP per output row has been computed. 

Although the FLOP per output row and the FLOP($C$) are precisely computed, we can still predict the FLOP($C$) in a symmetric way as how the reference design predicts the NNZ($C$) and establish the error analysis of the predicted FLOP($C$). Similar to the notations for the reference design, we denote the computed FLOP of the samples as $f^*$, the expected FLOP of the samples as $f$, the predicted FLOP($C$) as $F^*$, and the actual FLOP($C$) as $F$. The expected FLOP of the samples $f$  is defined as $f = pF$, where $p$ means the proportion of the matrix size of the sampled result matrix compared to the entire result matrix. We define the relative error of the $f^*$ to the expected $f$ as $\epsilon_f$. Then the entire prediction method and the error analysis for $F$ are shown in~\eqref{eq:F_star}.

\begin{equation}
    F^* = \frac{1}{p}f^* = \frac{1}{p}(1+\epsilon_f)f = (1+\epsilon_1)F.
    \label{eq:F_star}
\end{equation}

An intuition is that for the same sampled result matrix, if the computed $z^*$ is greater than the expected $z$, the computed $f^*$ will be more likely to be greater than the expected $f$ rather than smaller than the expected $f$. In other words, the relative error of  $z^*$ may have a positive correlation to the relative error of  $f^*$ for the same samples. 

Based on this intuition, we propose a novel prediction method as~\eqref{eq:Z2_star}. The proposed method first computes the precise FLOP and NNZ of the same samples (denoted as $f^*$ and $z_1^*$, respectively) so that the two variables may keep a certain degree of positive correlation. And then, the proposed method computes the predicted compression ratio as $r^* = f^*/z_1^*$. At last, the predicted NNZ($C$) is computed by dividing the actual FLOP($C$) by the predicted compression ratio. The output structure is easily computed by dividing the FLOP per output row by the predicted compression ratio. 

\begin{equation}
\begin{aligned}\label{eq:Z2_star}
    Z_2^* &= \frac{F}{r^*} = \frac{F}{f^*}z_1^* = \frac{F}{(1+\epsilon_f)f}(1+\epsilon_1)z \\
    & = \frac{1+\epsilon_1}{1+\epsilon_f}\frac{z}{p} = (1 + \frac{\epsilon_1 - \epsilon_f}{1+\epsilon_f})Z,
\end{aligned}
\end{equation}
where $Z_2^*$ is the predicted NNZ($C$) by the proposed method.

The relative error of the proposed method for predicting NNZ($C$) is:
\begin{equation}
    \epsilon_2 = \frac{\epsilon_1 - \epsilon_f}{1+\epsilon_f}.
    \label{eq:rela}
\end{equation} 

Based on~\eqref{eq:rela}, the relative error of the proposed method is close to the difference between $\epsilon_1$ and $\epsilon_f$ since the denominator is usually close to $1$. Therefore, the proposed method may achieve a good error neutralization between $\epsilon_1$ and $\epsilon_f$ if the two relative errors are positively correlated and close to each other. Specifically, if $\epsilon_1$ and $\epsilon_f$ are both positive values and $\epsilon_f \in [0, 2\epsilon_1]$, the relative error of $\epsilon_2$ will be smaller than $\epsilon_1$, which is our expectation.  Similarly, if $\epsilon_1$ and $\epsilon_f$ are both negative values and $\epsilon_f \in [2\epsilon_1, 0]$, the relative error of $\epsilon_2$ will also be ``smaller'' than $\epsilon_1$; the ``smaller'' here means more close to $0$. In a special case, if $\epsilon_f$ approaches $\epsilon_1$, $\epsilon_2$ will approach $0$.

\section{Parallel Implementation}\label{sec:impelmentation}
This section describes the implementations of two performance-critical tasks used by the proposed method and the reference design. 
The goal of the parallel implementations is to achieve low computing overheads. We target the implementation at the multi-core CPUs instead of GPUs since these computing tasks are relatively small and irregular, which may not efficiently utilize the massively parallel computing resources on GPUs. We implement the parallel algorithms with the OpenMP framework~\cite{openmp}.

\subsection{Computing the FLOP per Output Row}\label{sec:compute-flop}
The first performance-critical task is to compute the FLOP per output row.
Algorithm~\ref{alg:flop} shows the parallel implementation of this task, where the floprC denotes the actual FLOP per output row, and the total\_flop denotes the actual FLOP($C$). The FLOP per output row is the upper bounds of the number of nonzero elements per output row. The computation complexity of this algorithm is relatively small since only the row offsets and column indices of the $A$ matrix and the row offsets of the $B$ matrix are processed. We parallelize this algorithm by statically assigning the same number of rows to each CPU thread. The total\_flop (Line~\ref{line:critical}) is computed within a critical section provided by OpenMP~\cite{openmp}. 

\begin{algorithm}[h]
    \caption{Compute FLOP($C$)}
    \label{alg:flop}
    \hspace*{\algorithmicindent} \textbf{Input:} A.rpt, A.col, B.rpt, M. \\
    \hspace*{\algorithmicindent} \textbf{Output:} floprC, total\_flop.
    \begin{algorithmic}[1]
        \State{total\_flop = 0}
        \For{i = 0 \textbf{to} M-1 \textbf{in parallel}} 
        \State{local\_flop = 0}
        \For{j = A.rpt[i] \textbf{to} A.rpt[i+1]}
        \State{local\_flop += B.rpt[A.col[j]+1] - B.rpt[A.col[j]]}
        \EndFor
        \State{floprC[i] = local\_flop}
        \State{total\_flop += local\_flop}\label{line:critical} \Comment{computed in a critical section}
        \EndFor
    \end{algorithmic}
\end{algorithm}

\subsection{Computing the Predicted NNZ(C) by the Proposed Method}\label{sec:compute-nnz}
The second performance-critical computing task is to compute the NNZ of the sampled result matrix. Since the other computation steps are straightforward and with a low cost,  Algorithm~\ref{alg:estimate} directly shows the entire computation flow of how the proposed method computes the predicted NNZ($C$), which includes how the sampled NNZ is computed.

\begin{algorithm}
    \caption{Compute $Z_2^*$}
    \label{alg:estimate}
    \hspace*{\algorithmicindent} \textbf{Input:} A.rpt, A.col, B.rpt, B.col, floprC, total\_flop, M \\
    \hspace*{\algorithmicindent} \textbf{Output:} $Z_2^*$
    \begin{algorithmic}[1]
        \State{sample\_num = min(0.003 * M, 300)}\label{line:sample_num}
        \State{\textit{rand} = \textbf{new} \textbf{float} [sample\_num]} \label{line:alloc_rand}
        \State{Generate sample\_num random data in the range [0,1] and store them to the \textit{rand}} array\label{line:gen_rand}

        \State{Calculate the max floprC as the max hash table size: max\_tsize}\label{line:cal_tsize}
        
        \State{sample\_flop = 0, sample\_nnz = 0}
        \State{ht = \textbf{new int} [max\_tsize] for each CPU thread}\label{line:max_tsize}

        \For{r = 0 \textbf{to} sample\_num - 1 \textbf{in parallel}}\label{line:for_start} 
            \State{local\_nnz = 0}
            \State{rid = M * rand[r]}\label{line:select_row}
            \State{tsize = floprC[rid]} \label{line:tsize}
            \State{Initialize the front tsize elements in the \textbf{ht} array to $-1$}
            \For{i = A.rpt[rid] \textbf{to} A.rpt[rid+1]} \label{line:hash_start}
                \State{B\_row = B.col[i]}
                \For{j = B.rpt[B\_row] to B.rpt[B\_row + 1]}
                    \State{hash = (B.col[j] * HASH\_SCALE) \% tsize}
                    \While{true}
                        \If{ht[hash] == B.col[j]}
                        \State{\textbf{break}}
                        \ElsIf{ht[hash] == -1}
                        \State{local\_nnz += 1}
                        \State{ht[hash] = B.col[j]}
                        \State{\textbf{break}}
                        \Else
                        \State{hash = (hash + 1) \% tsize}
                        \EndIf
                    \EndWhile
                \EndFor
            \EndFor\label{line:hash_end}
            \State{sample\_nnz += local\_nnz} \Comment{computed in a critical section} \label{line:critical1} 
            \State{sample\_flop += floprC[rid]} \Comment{computed in a critical section}  \label{line:critical2}
        \EndFor\label{line:for_end}
        \State{$Z_2^*$ = total\_flop / sample\_flop * sample\_nnz} \label{line:estimate}
 
    \end{algorithmic}
\end{algorithm}

The method to compute the NNZ per output row is the same as the existing work~\cite{yusuke}, which utilizes the hash-based method. Line~\ref{line:hash_start} to Line~\ref{line:hash_end} in Algorithm~\ref{alg:estimate} show the hash-based method. The memory space of the hash table for each CPU thread is set as the largest floprC (Line~\ref{line:max_tsize}). However, the used memory space of the hash table for each output row is the FLOP of that row (Line~\ref{line:tsize}). Line~\ref{line:select_row} selects a random row from the $A$ matrix, where the random data is pre-computed and stored in an array named \textit{rand} (Line~\ref{line:gen_rand}). Line~\ref{line:for_start} is parallelized so that each CPU thread computes the same number of sampled rows. As a result of the parallelism, Line~\ref{line:critical1} and Line~\ref{line:critical2} are computed in a critical section~\cite{openmp}. In the end,  the predicted NNZ($C$) is easily computed as shown in Line~\ref{line:estimate}.

In the implementation of both the proposed method and the reference design, we empirically set the number of the sampled rows as $sample\_num = min(0.003M, 300)$, where $M$ is the number of rows of $A$. We set the maximum sampled rows as 300 to reduce the computing overheads when $M$ is large. The accuracy loss is often negligible since the prediction accuracy is often very high when $M$ is relatively large.

\section{Experiments}\label{sec:experiment}

\begin{table*}[h]
    \centering
    \caption{Detailed information of 25 matrices from the SuiteSparse datasets. CR represents the compression ratio.}
    \fontsize{9.5}{12}\selectfont
    \label{tab:matrix}
    \begin{tabular}{|l|l|l|l|l|l|l|l|l|}
        \hline
        \textbf{Id} & \textbf{Name} & \textbf{Rows} & \textbf{NNZ} & \textbf{NNZ/row} & \textbf{Max NNZ/row} & \textbf{FLOP of $A^2$} & \textbf{NNZ of $A^2$} & \textbf{CR of $A^2$} \\
        \hline
        1 & m133-b3 & \num{200200} & \num{800800} & 4.0 & \num{4} & \num{3203200} & \num{3182751} & 1.01 \\
        \hline
        2 & mac\_econ\_fwd500 & \num{206500} & \num{1273389} & 6.2 & \num{44} & \num{7556897} & \num{6704899} & 1.13 \\
        \hline
        3 & patents\_main & \num{240547} & \num{560943} & 2.3 & \num{206} & \num{2604790} & \num{2281308} & 1.14 \\
        \hline
        4 & webbase-1M & \num{1000005} & \num{3105536} & 3.1 & \num{4700} & \num{69524195} & \num{51111996} & 1.36 \\
        \hline
        5 & mc2depi & \num{525825} & \num{2100225} & 4.0 & \num{4} & \num{8391680} & \num{5245952} & 1.60 \\
        \hline
        6 & scircuit & \num{170998} & \num{958936} & 5.6 & \num{353} & \num{8676313} & \num{5222525} & 1.66 \\
        \hline
        7 & delaunay\_n24 & \num{16777216} & \num{100663202} & 6.0 & \num{26} & \num{633914372} & \num{347322258} & 1.83 \\
        \hline
        8 & mario002 & \num{389874} & \num{2101242} & 5.4 & \num{7} & \num{12829364} & \num{6449598} & 1.99 \\
        \hline
        9 & cage15 & \num{5154859} & \num{99199551} & 19.2 & \num{47} & \num{2078631615} & \num{929023247} & 2.24 \\
        \hline
        10 & cage12 & \num{130228} & \num{2032536} & 15.6 & \num{33} & \num{34610826} & \num{15231874} & 2.27 \\
        \hline
        11 & majorbasis & \num{160000} & \num{1750416} & 10.9 & \num{11} & \num{19178064} & \num{8243392} & 2.33 \\
        \hline
        12 & offshore & \num{259789} & \num{4242673} & 16.3 & \num{31} & \num{71342515} & \num{23356245} & 3.05 \\
        \hline
        13 & 2cubes\_sphere & \num{101492} & \num{1647264} & 16.2 & \num{31} & \num{27450606} & \num{8974526} & 3.06 \\
        \hline
        14 & poisson3Da & \num{13514} & \num{352762} & 26.1 & \num{110} & \num{11768678} & \num{2957530} & 3.98 \\
        \hline
        15 & filter3D & \num{106437} & \num{2707179} & 25.4 & \num{112} & \num{85957185} & \num{20161619} & 4.26 \\
        \hline
        16 & cop20k\_A & \num{121192} & \num{2624331} & 21.7 & \num{81} & \num{79883385} & \num{18705069} & 4.27 \\
        \hline
        17 & mono\_500Hz & \num{169410} & \num{5036288} & 29.7 & \num{719} & \num{204030968} & \num{41377964} & 4.93 \\
        \hline
        18 & conf5\_4-8x8-05 & \num{49152} & \num{1916928} & 39.0 & \num{39} & \num{74760192} & \num{10911744} & 6.85 \\
        \hline
        19 & cant & \num{62451} & \num{4007383} & 64.2 & \num{78} & \num{269486473} & \num{17440029} & 15.45 \\
        \hline
        20 & hood & \num{220542} & \num{10768436} & 48.8 & \num{77} & \num{562028138} & \num{34242180} & 16.41 \\
        \hline
        21 & consph & \num{83334} & \num{6010480} & 72.1 & \num{81} & \num{463845030} & \num{26539736} & 17.48 \\
        \hline
        22 & shipsec1 & \num{140874} & \num{7813404} & 55.5 & \num{102} & \num{450639288} & \num{24086412} & 18.71 \\
        \hline
        23 & pwtk & \num{217918} & \num{11634424} & 53.4 & \num{180} & \num{626054402} & \num{32772236} & 19.10 \\
        \hline
        24 & rma10 & \num{46835} & \num{2374001} & 50.7 & \num{145} & \num{156480259} & \num{7900917} & 19.81 \\
        \hline
        25 & pdb1HYS & \num{36417} & \num{4344765} & 119.3 & \num{204} & \num{555322659} & \num{19594581} & 28.34 \\
        \hline
    \end{tabular}
\end{table*}

In this section, we compare the accuracy of the predicted NNZ($C$) of the proposed method and the reference design. We also show the prediction overheads of the proposed method compared with a state-of-the-art SpGEMM library BRMerge-Precise~\cite{brmerge}. 

\subsection{Predicting Accuracy}\label{sec:eva-accuracy}
Recall that for both the proposed method and the reference design, the output structure is computed by dividing the FLOP per output row by the predicted compression ratio, where the predicted compression ratio is computed by dividing the total FLOP  by the predicted total NNZ. Since the FLOP per output row and the total FLOP are precisely computed, the accuracy of the predicted output structure, the predicted compression ratio, and the predicted NNZ($C$) can be seen as equivalent. Therefore, we only compare the prediction accuracy of the predicted NNZ($C$).

For the diversity of the evaluation, we select 25 representative real-world sparse matrices from the SuiteSparse matrix collection~\cite{suitesparse}. Table~\ref{tab:matrix} shows the detailed information of the 25 sparse matrices. We try to multiply these 25 matrices with each other to obtain more test cases with various sparse structures and matrix dimensions. One problem is that the two input matrices may not be multiplied due to mismatched matrix dimensions. To tackle this problem, we reshape either the first or the second input matrix. For example, if the dimensions of the two input matrices are 10 $\times$ 10 and 5 $\times$ 5, we reshape the first matrix to a 10 $\times$ 5 matrix by keeping its left 5 columns. If the dimensions of the two input matrices are 5 $\times$ 5 and 10 $\times$ 10, we reshape the second matrix to a 5 $\times$ 10 matrix by keeping its top 5 rows. As a result, we construct 625 matrix multiplication test cases to evaluate the accuracy.

We compare the three relative errors: $\epsilon_1 = (Z_1^* - Z)/Z$, $\epsilon_f =  (F^* - F)/F$, and $\epsilon_2 =  (Z_2^* - Z)/Z$, which are described in section~\ref{sec:method}. Recall that $F$ can be precisely computed with low costs, which means we do not need to predict the total FLOP($C$) in real cases. We show the relative prediction error of $\epsilon_f$ compared to $\epsilon_1$ to observe if they have the expected positive correlation.

Experiments on the 625 test cases show that the average absolute relative errors of $\epsilon_1$, $\epsilon_f$, and $\epsilon_2$ are 8.12\%, 8.59\%, and 1.56\%, respectively. Whereas the worst absolute relative errors of $\epsilon_1$, $\epsilon_f$, and $\epsilon_2$ are 158\%, 155\%, and 25\%, respectively. Moreover, the proposed method is more accurate than the reference design on 81.4\% of the 625 test cases.
The overall results show that the proposed method is much more accurate than the reference design. Considering the difference between the proposed method and the reference design, we can infer that the relative errors of the sampled FLOP and sampled NNZ are positively correlated and close to each other in most scenarios. Therefore, the proposed method can achieve good error neutralization  by dividing the sampled FLOP by the sampled NNZ (see section~\ref{sec:proposed-method}).


\begin{table*}[h]
	\centering
	\fontsize{9.5}{12}\selectfont
    \caption{The relative errors of 20 representative test cases.}
    \label{tab:rela}
	\begin{tabular}{|l|l|l|l|l|l|l|l|l|}
	\hline
	  & A   & B  & {smaple\_num} & CR  & NNZ(C) & $\epsilon_1$(\%) & $\epsilon_f$(\%) & $\epsilon_2$(\%)  \\ 
	\hline
	1  & 2cubes\_sphere & consph            & 300         & 1.5  & 64800734 & -2.31    & -2.6     & 0.29     \\ \hline
	2  & cage12         & patents\_main     & 300         & 1    & 4611949  & 3.61     & 3.62     & -0.01    \\ \hline
	3  & cage15         & majorbasis        & 300         & 1.1  & 15990225 & 16.38    & 15.73    & 0.56     \\ \hline
	4  & delaunay\_n24  & mario002          & 300         & 1    & 12553686 & -22.49   & -22.81   & 0.42     \\ \hline
	5  & delaunay\_n24  & cop20k\_A         & 300         & 1.01 & 15604104 & -45.88   & -46.32   & 0.81     \\ \hline
	6  & m133-b3        & rma10             & 300         & 1.14 & 8336596  & 4.41     & 5.57     & -1.11    \\ \hline
	7  & majorbasis     & 2cubes\_sphere    & 300         & 1.08 & 22688054 & -0.49    & -0.5     & 0.01     \\ \hline
	8  & mario002       & webbase-1M        & 300         & 1.21 & 6866846  & -11.98   & -16.68   & 5.65     \\ \hline
	9  & mc2depi        & poisson3Da        & 300         & 1.02 & 1366481  & 66.23    & 70.54    & -2.53    \\ \hline
	10 & pwtk           & consph            & 300         & 5.99 & 54168970 & -11.32   & -9.81    & -1.68    \\ \hline
	11 & shipsec1       & rma10             & 300         & 4.67 & 27713808 & -7.48    & -5.53    & -2.07    \\ \hline
	12 & scircuit       & poisson3Da        & 300         & 1.04 & 1848459  & 8.76     & 8.06     & 0.65     \\ \hline
	13 & scircuit       & mac\_econ\_fwd500 & 300         & 1.11 & 5313337  & 1.44     & 1.36     & 0.08     \\ \hline
	14 & rma10          & pdb1HYS           & 140         & 8.3  & 23240867 & -3.07    & -2.43    & -0.66    \\ \hline
	15 & pwtk           & shipsec1          & 300         & 5.41 & 77530890 & -3.59    & -4.29    & 0.73     \\ \hline
	16 & cage12         & hood              & 300         & 1.23 & 83406736 & -0.31    & 0.2      & -0.51    \\ \hline
	17 & 2cubes\_sphere & cant              & 300         & 1.62 & 40235181 & -0.19    & -3.47    & 3.4      \\ \hline
	18 & rma10          & offshore          & 140         & 1.53 & 25255211 & -0.01    & 0.28     & -0.29    \\ \hline
	19 & filter3D       & filter3D          & 300         & 4.26 & 20161619 & 1.74     & 4.47     & -2.62    \\ \hline
	20 & hood           & poisson3Da        & 300         & 1.12 & 17777942 & -0.39    & 0.98     & -1.35    \\ \hline
\end{tabular}
\end{table*}

Moreover, the correlation coefficient~\cite{corre} of $\epsilon_1$ and $\epsilon_f$ on the 625 test cases is 97.01\%, which statistically shows a strong positive correlation between the sampled FLOP and sampled NNZ of the same randomly selected samples.

To show the better accuracy of the proposed method more intuitively, we show the relative errors of $\epsilon_1$, $\epsilon_f$, and $\epsilon_2$ on 20 representative test cases in Table~\ref{tab:rela}. For most test cases, the prediction accuracy of the proposed method is much more accurate than the reference design. For example, the fifth test case in Table~\ref{tab:rela} shows that the relative error of predicting the NNZ($C$) and FLOP($C$) by only using the NNZ or FLOP of the samples are -45.88 and -46.32, respectively. In contrast, the relative error of predicting NNZ($C$) by the proposed method is only 0.81. This test case shows significant error neutralization between the sampled FLOP and sampled NNZ. Also note that the relative errors of $\epsilon_1$, $\epsilon_f$, and $\epsilon_2$ of all the test cases exactly meet the equation described by~\eqref{eq:rela}.

The last five test cases in Table~\ref{tab:rela} show that the signs of  $\epsilon_1$ and $\epsilon_f$ are different. We observe that when the signs of  $\epsilon_1$ and $\epsilon_f$ are different, the absolute values of $\epsilon_1$ and $\epsilon_f$ are usually close to zero. As a result, the relative error of predicting NNZ($C$) by the proposed method is also relatively small in such cases.

\subsection{Computing Overhead}\label{sec:performance}

Two performance-critical tasks in the proposed method and the reference design are computing the FLOP per output row (denoted as computing FLOP) and computing the NNZ of the samples (see section~\ref{sec:compute-flop} and section~\ref{sec:compute-nnz}). Since the computation complexity of computing the sampled NNZ is similar to predicting the NNZ($C$) by the proposed method after the FLOP per output row has been computed, we directly show the computing overheads of predicting NNZ($C$) by the proposed method (denoted as predicting $Z_2^*$).

\begin{figure*}[h]
    \centering
    \includegraphics[width=0.9\textwidth]{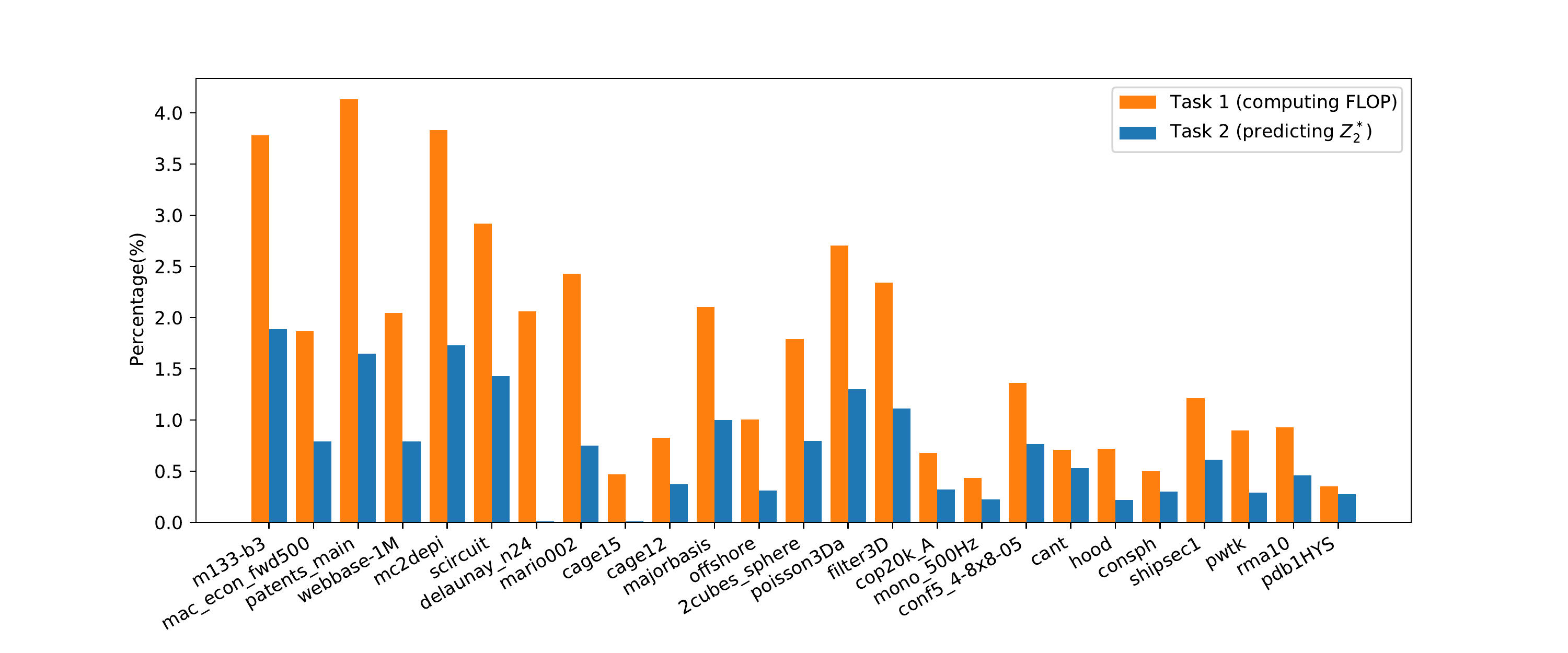}
    \caption{Relative execution time of two performance-critical algorithms compared to BRMerge-Precise.}
    \label{fig:estimate_perf}
\end{figure*}

We compare the relative execution time of the two tasks compared to a state-of-the-art SpGEMM library BRMerge-Precise~\cite{brmerge}. The execution time is measured as the average execution time of ten runs after one warm-up run. Fig.~\ref{fig:estimate_perf} shows the relative execution time in percentage on the matrix square benchmark with the 25 sparse matrix in Table~\ref{tab:matrix}. 
The average relative execution time of computing FLOP and predicting $Z_2^*$ compared to BRMerge-Precise are 1.68\% (up to 4.12\%) and 0.72\% (up to 1.89\%), respectively, which is a relatively small computing overhead.

In most SpGEMM algorithms, computing the FLOP per output row is a necessary task for either the upper-bound allocation method~\cite{bhsparse, ESC, ESC2, yusuke} or the load balance of the symbolic phase~\cite{nsparse, speck}. Therefore, the actual computing overheads of the proposed prediction method only take an average of 0.72\% (up to 1.89\%) execution time compared to the state-of-the-art SpGEMM library BRMerge-Precise, which is a negligible cost.

\section{Conclusion}\label{sec:conclude}
Computing or predicting the output structure of SpGEMM is an important task for efficient memory allocation and load balance of SpGEMM, which greatly impacts the overall performance. In this paper, we propose a novel sampling-based prediction method that utilizes the positive correlation between the sampled FLOP and sampled NNZ of the same samples to achieve a certain degree of error neutralization. The proposed method achieves much more accurate prediction accuracy than the reference design of the existing sampling-based method. For low computing overheads, the proposed method adopts the row-wise dataflow and only samples up to 0.003 of the total rows of the first input matrix. We also propose the parallel implementation of the proposed method targeting the multi-core CPUs. The computing overheads of the proposed method only take on average 0.72\% execution time compared to the overall execution time of a state-of-the-art SpGEMM library BRMerge-Precise.

\bibliographystyle{ieeetr}
\bibliography{cite.bib}

\end{document}